\documentclass[PRA,superscriptaddress,longbibliography,nofootinbib,twocolumn]{revtex4-2}
\usepackage[main=english]{babel}

\usepackage{layouts}

\usepackage{graphicx} 
\usepackage[]{xcolor}

\usepackage{amssymb}
\usepackage{amsmath}
\usepackage{amsthm}
\usepackage{mathtools}

\usepackage{bm}
\usepackage{bbold} 
\usepackage{physics}
\usepackage[version=4]{mhchem}
\usepackage{siunitx}

\definecolor{LEI-blue}{cmyk}{1,.75,0,.35} 
\definecolor{LEI-orange}{cmyk}{0,.62,.97,0} 
\definecolor{niceblue}{rgb}{.1, .25, .8}

\usepackage[colorlinks=true, linkcolor=niceblue, citecolor=niceblue]{hyperref}
\usepackage{orcidlink}

\newcommand{\dens}{\boldsymbol{\rho}}
\newcommand{\pot}{\boldsymbol{\mu}}
\newcommand{\gs}{\text{GS}}
\newcommand{\ml}{\text{ML}}
\newcommand{\EUT}{F} 

\newcommand{\aQa}{\affiliation{%
    $\langle a Q a^L \rangle$ Applied Quantum Algorithms,
    Universiteit Leiden, The Netherlands}
}

\begin{document}

\title{Learning Minimal Representations of Fermionic Ground States}

\author{Felix Frohnert \orcidlink{0000-0003-3717-6352}}
\email[E-mail:]{f.frohnert@liacs.leidenuniv.nl}
\aQa

\author{Emiel Koridon \orcidlink{0000-0002-7712-7638}}
\aQa

\author{Stefano Polla \orcidlink{0000-0003-3909-0448}}
\email[E-mail:]{polla@lorentz.leidenuniv.nl}
\aQa

\date{December 2025}

\begin{abstract}
We introduce an unsupervised machine-learning framework that discovers optimally compressed representations of quantum many-body ground states.
Using an autoencoder neural network architecture on data from $L$-site Fermi-Hubbard models, we identify minimal latent spaces with a sharp reconstruction quality threshold at $L-1$ latent dimensions, matching the system's intrinsic degrees of freedom.
We demonstrate the use of the trained decoder as a differentiable variational ansatz to minimize energy directly within the latent space. 
Crucially, this approach circumvents the $N$-representability problem, as the learned manifold implicitly restricts the optimization to physically valid quantum states.

\end{abstract}

\maketitle

\section{Introduction}\label{sec:introduction}
The exponential growth of the Hilbert space with system size constitutes a fundamental challenge in quantum many-body physics, rendering direct simulation intractable beyond modest particle numbers~\cite{feynman_simulating_1982}. 
Despite this barrier, approximate methods enable accurate simulations of systems containing tens to hundreds of particles across quantum chemistry and condensed matter physics~\cite{tindall_efficient_2024,patra_efficient_2024,menczer_parallel_2024,brabec_massively_2020,nagy_approaching_2019}.
These successes rely on exploiting the structure of the underlying physical models through tailored representations of the relevant quantum states, which compress the information of the exponentially large Hilbert space into a manageable form while remaining useful for specific tasks.

Modern electronic-structure methods exemplify this principle through diverse approaches~\cite{schollwoeck_density-matrix_2011,chan_density_2011,orus_tensor_2019,helgaker2000molecular,sherrill_frontiers_2010}. 
Tensor networks leverage entanglement area laws to identify a relevant state manifold within the Hilbert space, enabling algorithms such as the density matrix renormalization group (DMRG)~\cite{white_density_1992,white_density-matrix_1993,verstraete_matrix_2008,eisert_colloquium_2010}. 
Density functional theory (DFT) reformulates the many-body problem in terms of the electron density; whose dimensionality scales linearly with the system size, at the cost of approximate exchange-correlation functionals~\cite{burkeDFT2013,burke_perspective_2012,mardirossian_thirty_2017}. 
Coupled cluster methods construct size-extensive wave functions through systematic excitations from mean-field references~\cite{bartlett_coupled-cluster_2007,crawford2007introduction,zhang_coupled_2019}. 
Each approach exploits specific features, locality of interactions, symmetries, or correlation patterns that render compression possible. 
Indeed, for arbitrary Hamiltonians where any state can be a ground state, no universal compression scheme can exist.

These hand-crafted representations, while powerful, are inherently limited by their design principles. 
Tensor networks excel for states with mostly short-range entanglement, DFT on weakly correlated systems, and coupled cluster methods for capturing dynamic correlation. 
Strongly correlated systems, characterized by competing interactions that preclude simple effective descriptions, remain particularly challenging~\cite{georges_strongly_2004}. 
This limitation motivates a fundamental question: 
can compressed representations be learned directly from data, adapting to the specific system and task of interest rather than relying on predetermined structures?

Recent advances in machine learning offer a natural route toward this goal~\cite{dawid_modern_2025,rocchetto_learning_2018,dunjko_machine_2018,carleo_machine_2019,zang_machine_2024, sager-smith_reducing_2022,carleo_solving_2017,sajjan_quantum_2022}.
Representation learning techniques aim to automatically extract informative descriptors from data, transforming information into formats that are optimal for downstream objectives, such as prediction, interpretability, or transferability~\cite{bengio_representation_2014,alshammari_i-con_2025,chen_simple_2020, costaSolving2024}.
Autoencoding frameworks are an example of this, as they discover compact encodings by mapping their input through a bottleneck and then reconstructing it, revealing underlying factors of variation while discarding redundancy in the data~\cite{prince_understanding_2023, chen_auto-encoders_2023, moller_learning_2025,frohnert_explainable_2024}.

In this Letter, we introduce an unsupervised machine learning framework that learns minimal compressed representations of quantum many-body ground states, optimized for variational energy minimization in the latent space.
The model automatically discovers the essential degrees of freedom of the family of Hamiltonians on which it is trained, without prior knowledge of the underlying physics.
We apply this framework to the Fermi–Hubbard model in the strongly correlated regime and investigate the natural compression limit and its relation to the system’s intrinsic degrees of freedom~\cite{hubbard_electron_1997, arovas_hubbard_2022,cheuk_observation_2016}.
The learned representations enable determining the ground state energies of new Hamiltonian instances through variational energy optimization directly in the latent space; crucially, this is achieved without the need to explicitly enforce $N$-representability constraints.

This work bridges two complementary perspectives: the physics-driven identification of efficient quantum state descriptors and the data-driven discovery of task-specific representations through machine learning.
The resulting framework demonstrates that learned compressions can serve both as practical variational ansätze for many-body calculations and as diagnostic tools for complexity in ground state physics.
While the Fermi-Hubbard model provides a controlled testbed in which the degrees of freedom are known analytically, the broader goal is to develop a methodology applicable to complex quantum systems, such as molecular, material, or experimental many-body settings, where the relevant effective variables are not known a priori.

\begin{figure}[t]
    \centering
    \includegraphics[width=\linewidth]{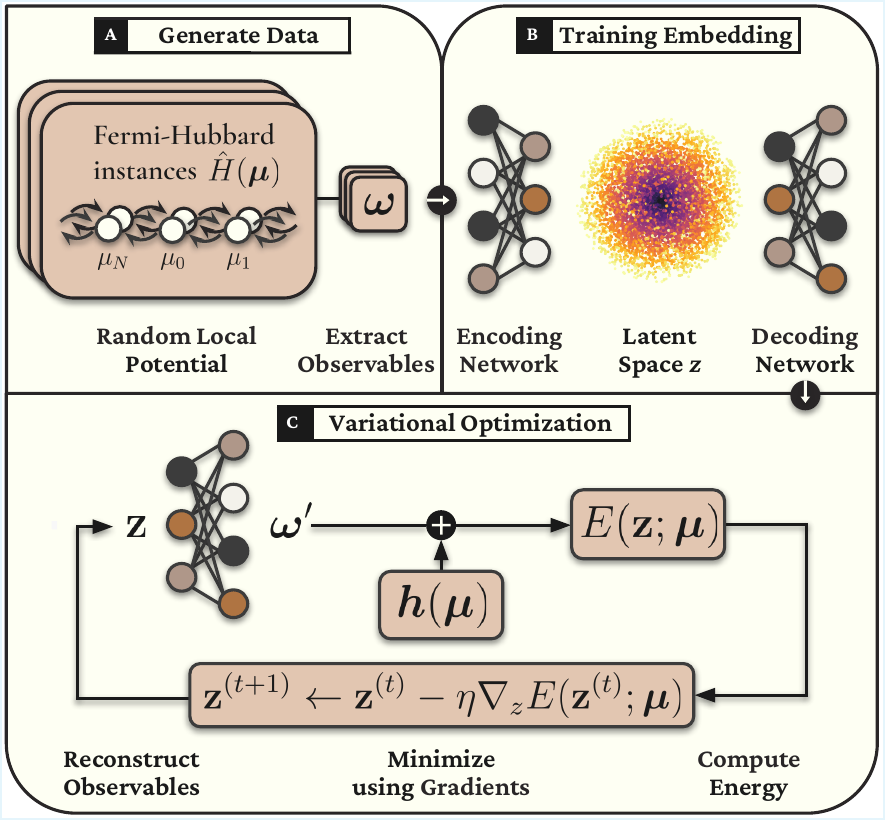}
    \caption{ \textbf{Overview}: workflow for learning minimal representations of fermionic ground states.
A) The process starts by generating instances of the Hubbard Hamiltonian $\hat H(\pot)$ from random local potentials $\pot$.
For each instance, the expectation values of the Hamiltonian terms $\boldsymbol{\omega}$ are computed via exact diagonalization.
B)
The resulting dataset is used to train a neural network–based autoencoder that compresses $\boldsymbol{\omega}$ into a low-dimensional latent representation $\mathbf{z}$.
C) The decoder defines a differentiable mapping from latent space to the expectation values of Hamiltonian terms, and consequently to physical observables such as the energy.
This property enables energy optimization directly in the latent space for new, unseen potentials. 
    } 
    \label{fig:overview}
\end{figure}

\section{Methods}
The overarching goal of our machine learning framework is to learn compressed representations of ground states for a given family of Hamiltonians that can serve as variational ansätze, as illustrated in Fig.~\ref{fig:overview}.

\noindent \textbf{Dataset and task} ---
We consider the family of one-dimensional Fermi–Hubbard models on $L$ sites with $N$ electrons, parameterized by the site-dependent chemical potential $\boldsymbol{\mu}=(\mu_1, ..., \mu_L)$:
\begin{align}\label{eq:hubbardHam}
    \hat H (\boldsymbol{\mu})= & -t \sum_{i}^L\sum_{\sigma}\big(\hat c^\dagger_{i\sigma}\hat c^{}_{i+1\sigma}+\text{h.c.}\big) \\
    &+U \sum_{i=1}^L \hat n_{i\uparrow}\hat n_{i\downarrow}
    + \sum_{i=1}^L \mu_i \sum_{\sigma} \hat n_{i\sigma}, \nonumber
\end{align}
with fixed hopping $t=1$ and interaction strength {$U=4$}.
This matches the strongly-correlated regime, which is especially challenging to solve on $d\geq2$-dimensional lattices~\cite{leblancSolutions2015}.
We restrict the analysis to the block with fixed particle number $\sum_{i, \sigma}{\hat{n}_{i,\sigma}} = N$ (choosing $N$ at or close to half filling, $N=L/2$), zero spin projection $S_z = 0$, and total spin singlet $\boldsymbol{S}^2 = 0$.
We limit ourselves to system sizes $L\leq14$ that can be solved by exact diagonalization to ensure high-quality data.

The dataset for our representation learning task is generated by sampling random potentials $\boldsymbol{\mu}$ from the distribution described in Ref.~\cite{koridon_learning_2025} (see App.~\ref{appx:dataset}). 
For each instance, we compute the ground state $\ket{\psi(\boldsymbol{\mu})}$ via exact diagonalization and extract the vector of Hamiltonian expectation values:
\begin{equation}
    \boldsymbol{\omega} = \Big\{\big( \langle \sum_\sigma \hat c^\dag_{i\sigma} \hat c_{i+1,\sigma} \rangle, 
    \langle\sum_\sigma\hat{n}_{i\sigma}\rangle, 
    \langle\hat{n}_{i\uparrow}\hat{n}_{i\downarrow}\rangle \big) \Big\}_{i=1,...,L}.
\end{equation}
This representation provides a compact description of the underlying states in which the total energy is recovered exactly via the contraction of $\boldsymbol{\omega}$ with the Hamiltonian coefficients $\boldsymbol{h}(\boldsymbol{\mu}) := \{(t, U, \mu_i)\}_{i=1,\dots,L}$:
\begin{equation}
\langle \hat H (\boldsymbol{\mu}) \rangle
= \boldsymbol{h}(\boldsymbol{\mu}) \cdot \boldsymbol{\omega}.
\end{equation}
Crucially, unlike the exponentially-large wavefunction, the dimensionality of $\boldsymbol{\omega}$ scales linearly with $L$.

The motivation for using this representation as the starting point for our machine-learning compression framework is twofold.
First, while $\boldsymbol{\omega}$ constitutes the minimal representation required to exactly extract the energy, the dimensionality of the underlying parameters generating the data is strictly smaller than the dimensionality of the Hamiltonian term description ($\dim(\boldsymbol{\mu}) < \dim(\boldsymbol{\omega})$), hinting towards additional compressibility.
Second, the target application for our models is variational energy optimization; 
yet we cannot optimize within the space of $\boldsymbol{\omega}$ directly. 
Direct variational minimization is fundamentally precluded by the $N$-representability problem: arbitrary vectors in $\boldsymbol{\omega}$-space do not necessarily map to physical quantum states. 
Our compression scheme aims to automatically identify a manifold in $\boldsymbol{\omega}$-space that implicitly enforces these physical constraints.

Our methodology is not limited to Hamiltonian term expectation values; 
we explore the compression of another meaningful set of observables, the spin-adapted two-electron reduced density matrix (2-RDM), in App.~\ref{app:2rdm}.
The 2-RDM encodes all two-body correlation functions and serves as the fundamental variable in variational 2-RDM~\cite{coleman_structure_1963,lowdin1955quantum, delgado-granados_machine_2025} methods. 
The 2-RDM provides a richer, though higher-dimensional, representation of the ground states from which all two-particle observables (including the energy) can be computed.

\noindent \textbf{Architecture} ---
We train neural-network autoencoders~\cite{prince_understanding_2023,chen_auto-encoders_2023} to learn compressed representations of Hubbard model ground states.
Below, we outline the structure and training of the best-performing models; 
the motivation behind each design choice is discussed in the ablation study (App.~\ref{apx:ablation}).
The models comprise an encoder network, $E_{\boldsymbol{\theta}}: \boldsymbol{\omega} \mapsto \mathbf{z}$, and a decoder network, $D_{\boldsymbol{\phi}}: \mathbf{z} \mapsto {\boldsymbol{\omega}'}$, where $\boldsymbol{\omega}$ denotes the input observables and $\mathbf{z} \in \mathbb{R}^d$ the latent representation, in which the latent space dimensionality $d$ is a hyperparameter of the model.
The encoder compresses the input through four non-linear fully connected layers, reducing its initial dimension (set by the input size) to a $d$-dimensional latent bottleneck.
The decoder mirrors this structure, reconstructing the observables from the compressed representation.

The model is trained to minimize the mean-squared reconstruction loss
\begin{equation}
    \mathcal{L}_{\mathrm{rec}}=\|\boldsymbol{\omega}-D_{\boldsymbol \phi}(E_\theta(\boldsymbol{\omega}))\|_2^2.
\end{equation}
The data is standardized to ensure numerical stability during training.
We employ Optuna~\cite{akiba_optuna_2019} to guide hyperparameter optimization, tuning both architectural choices and training parameters. 
All architectural and training details are provided in App.~\ref{appx:network}.

\noindent \textbf{Regularization} ---
To ensure stable training and learned latent representations that are geometrically well-behaved, we augment the reconstruction loss with four regularization terms.

First, we employ a radial well loss that bounds the latent norm to a finite radius, preventing unbounded encodings~\cite{prince_understanding_2023}. 
This constraint helps maintain stable representations by limiting the magnitude of latent vectors:
\begin{equation}
    \mathcal{L}_{\text{well}}=\max\!\bigl(0,\|\mathbf z\|_2-r\bigr)^2, \label{eq:well}
\end{equation}
where $r$ denotes the radius parameter that defines the maximum allowed norm in latent space.

Second, we incorporate a contrastive repulsion term to preserve the relative distances between distinct ground states in latent space, preventing degeneracy or collapse of the representation. 
This regularization follows principles inspired by contrastive learning methods~\cite{chen_simple_2020}, ensuring that distinct inputs maintain meaningful separation in the learned representation. 
For a batch $\{( \boldsymbol{\omega}_i,\mathbf z_i)\}$, the repulsion loss is formulated as:
\begin{equation}
    \mathcal{L}_{\text{repel}}
=\sum_{i<j}
\frac{\|\boldsymbol{\omega}_i-\boldsymbol{\omega}_j\|_2^2}{\|\mathbf z_i-\mathbf z_j\|_2^2+\varepsilon}
\;\Big/\;
\sum_{i<j}
\|\boldsymbol{\omega}_i-\boldsymbol{\omega}_j\|_2^2, \label{eq:repulsion}
\end{equation}
where $\varepsilon$ is a small constant for numerical stability.

Finally, we apply Lipschitz regularization to the encoder and decoder weights to promote smoothness and improve training stability. 
This approach constrains the Lipschitz constant of each layer, limiting how rapidly the network output can change with respect to its inputs. 
Conceptually, this has an effect similar to weight decay, since both penalize large weight magnitudes; however, Lipschitz regularization more directly controls the global smoothness of the learned mapping. 
Empirically, this encourages smoother latent-to-output interpolations and stabilizes gradients during optimization. 
Following a variant of the formulation in Ref.~\cite{liu_learning_2022}, we implement per-layer trainable Lipschitz log-bounds, which are softly enforced via a differentiable constraint:
\begin{equation}
\mathcal{L}_{\text{lip}} = \sum_i \log\operatorname{softplus}(c_i),
\label{eq:lip}
\end{equation}
where $c_i$ denotes a trainable log-bound on the $\infty$-norm of the weight matrix $\hat{W}_i$ for layer $i$. 
Each layer’s weight matrix $\hat{W}_i$ is defined by rescaling the trainable parameter matrix $W_i$ according to this bound,
\begin{equation}
\hat W_i=\mathrm{diag}\!\left(\min\!\Big[1,\tfrac{\mathrm{softplus}(c_i)}{\sum_k |(W_i)_{rk}|}\Big]_{r}\right) W_i,
\end{equation}
so that $\operatorname{softplus}(c_i)$ defines an adaptive upper limit on the layer’s Lipschitz constant.
The softplus function ensures positivity while maintaining smooth gradients, allowing each layer to adaptively learn its Lipschitz limit. 
This form of regularization yields a decoder that is both smoother and more robust than standard weight-decayed networks \cite{liu_learning_2022}.

The total training objective combines and weights all terms,
\begin{equation}
    \mathcal{L}
=
\mathcal{L}_{\mathrm{rec}}
+\alpha\,\mathcal{L}_{\text{well}}
+\beta\,\mathcal{L}_{\text{repel}}
+\gamma\,\mathcal{L}^{\text{enc}}_{\text{lip}}+\delta\,\mathcal{L}^{\text{dec}}_{\text{lip}},
\end{equation}
where $\mathcal{L}_\text{lip}^\text{enc}$ and $\mathcal{L}_\text{lip}^\text{dec}$ are the Lipschitz losses Eq.~\eqref{eq:lip} separately computed on the encoder and decoder layers, respectively.
This aims to learn a bounded, uniformly covered, and smooth latent manifold that encodes meaningful physical properties. 
The ablation study in App.~\ref{apx:ablation} illustrates how each regularization term affects this goal.

\noindent \textbf{Energy optimization} --- After training, the decoder $ D_{\boldsymbol{\phi}} $ defines a differentiable mapping from the latent representation $ \mathbf{z} $ to the corresponding Hamiltonian terms or observables. 
For a given potential $ \boldsymbol{\mu} $, the total energy can then be expressed as
\begin{equation}
E_\phi(\mathbf{z}; \boldsymbol{\mu}) = \boldsymbol{h}(\boldsymbol{\mu}) \cdot D_{\boldsymbol{\phi}}(\mathbf{z}), \label{eq:energy}
\end{equation}
where $ \boldsymbol{h}(\boldsymbol{\mu}) $ denotes the potential-dependent coefficients of the Hamiltonian, and $D_{\boldsymbol{\phi}}(\mathbf{z}) $ outputs the corresponding expectation values.
This formulation yields a fully differentiable map from latent space to energy, enabling the direct optimization of the latent variable to identify the ground state representation for a given potential:
\begin{equation}
\mathbf{z}^\star = \arg\min_{\mathbf{z}} E_\phi(\mathbf{z}; \boldsymbol{\mu}). \label{eq:opt-latent}
\end{equation}
The optimization can be performed via gradient descent while keeping the decoder parameters fixed, using gradients obtained through automatic differentiation,
$\mathbf{z}^{(t+1)} \leftarrow \mathbf{z}^{(t)} - \eta \, \nabla_{\mathbf{z}} E_\phi(\mathbf{z}^{(t)}; \boldsymbol{\mu})$.
This allows the latent space itself to serve as a search domain for new ground state configurations under varying external conditions.

We emphasize that this is not a trivial task. 
One might, in principle, attempt to minimize the energy directly with respect to the Hamiltonian expectation values $\boldsymbol{\omega}$, but this is fundamentally obstructed by the $N$-representability problem: 
not every collection of expectation values corresponds to a physical $N$-electron wavefunction. 
Enforcing the required $N$-representability constraints would entail imposing a hierarchy of semidefinite conditions, the computational cost of which grows prohibitively with the system sizes~\cite{lowdin1955quantum,nakata2008variational,mazziotti2002variationalb,delgado-granados_machine_2025}.
Our approach of optimizing in the learned latent space operates independently of these explicit conditions. 
Although the decoder is not explicitly constrained to generate $N$-representable states, a high quality of reconstruction and a sufficiently regular latent space can result in the optimization naturally remaining within or close to the physical ground state manifold.

\section{Results}\label{sec:results}
\noindent \textbf{Learning the Intrinsic Structure of the Ground State Manifold} ---
\begin{figure}[b]
    \centering
    \includegraphics[width=\columnwidth]{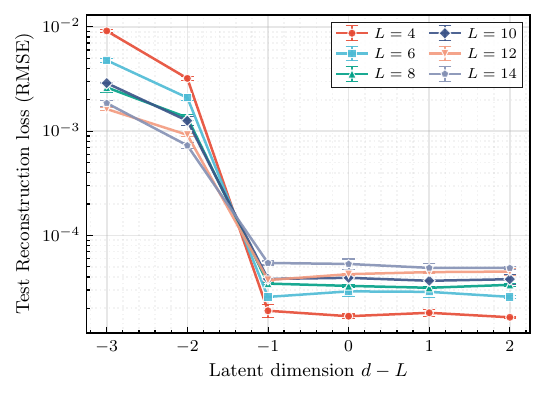}
    \caption{\textbf{Compression threshold}:
        test reconstruction loss (RMSE) of Hamiltonian-term vectors as a function of latent dimension $d$. Data points represent the mean and standard deviation across three independent models. A sharp decrease in error is observed at $d = L - 1$, identifying a compression threshold that coincides with the intrinsic number of independent degrees of freedom in the half-filled Hubbard ground state. This transition marks the point at which the autoencoder fully captures the physically relevant subspace of Hamiltonian observables.
    }
    \label{fig:compression}
\end{figure}
We first train and test the specified autoencoders on a set $2L\times 10^5$ of Hamiltonian term expectation values $\boldsymbol \omega$ for even system sizes ${L = [4, \ldots, 14]}$ at half-filling ${N=L/2}$.
We train the models for latent dimensions ${d \in [L-3, \ldots, L+2]}$, while keeping all other architectural hyperparameters fixed. 
This controlled scaling isolates the effect of the information bottleneck size on the representational capacity of the model.

Fig.~\ref{fig:compression} visualizes the test reconstruction loss as a function of latent dimension $d$.
We plot the mean and standard deviation over three independent models with random initialization.
Across all system sizes, the loss decreases with increasing $d$, exhibiting a sharp drop at $d = L - 1$, followed by saturation for $d \ge L - 1$.
This sudden improvement marks a compression threshold: for $d<L-1$ the latent space is insufficient to capture all relevant physical variations, while additional dimensions beyond $L-1$ provide no further benefit.

This behavior mirrors the physical structure of the underlying Hamiltonian family:
the number of degrees of freedom in the definition of $\hat H (\pot)$ corresponds to $L$, i.e.~the dimension of $\pot$.
However, as the particle number is fixed, a constant shift in the potential $\mu_i \mapsto \mu_i + c$ will shift the Hamiltonian by a constant $c\times N$ without affecting the ground state.
Thus, the number of independent degrees of freedom for the ground states of this Hamiltonian family is $L-1$. 
Empirically, the autoencoder autonomously identifies the true number of independent degrees of freedom, discovering through unsupervised compression that the ground state manifold is intrinsically $(L-1)$ dimensional.
The transition at $d = L - 1$ reveals a direct correspondence between empirically-optimal compression and the physical structure of the Hubbard model, 
demonstrating that at this point, the learned latent representation effectively captures the topology of the ground state family covered by the training distribution with minimal redundancy.
In App.~\ref{appx:odd}, we learn optimal compressed representations of systems with an uneven system size $L$ at $N=L/2 \pm 1$, to showcase how degeneracies in the underlying data affect the quality of reconstruction and latent manifold. 
\newline

\noindent \textbf{Latent Geometry of the Ground State Representations} ---
\begin{figure}[t]
    \centering
    \includegraphics[width=\columnwidth]{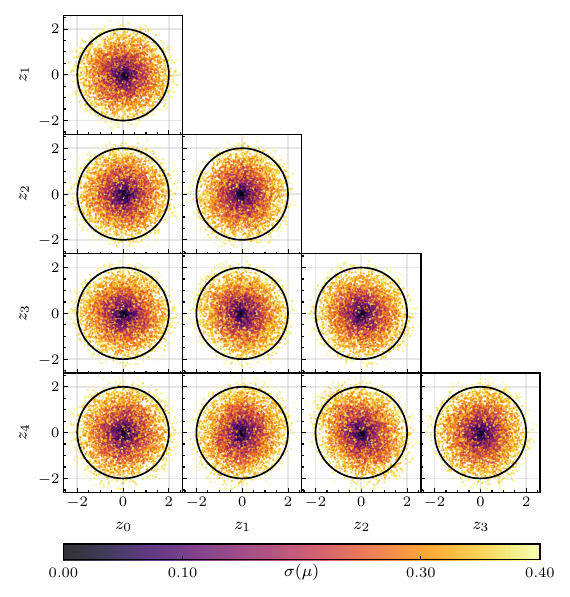}
    \caption{\textbf{Visualizing the learned representation}:
        latent space of the autoencoder for $(L = N = 6)$, color-coded by the strength of the potential $\boldsymbol{\mu}$.
        The circular structure in the feature-wise projections results from the well constraint and uniform coverage enforced by contrastive repulsion, while the smooth radial color gradient shows that states with similar potential strengths cluster together, indicating that the latent geometry captures physically meaningful structure in the ground state manifold.
    }
    \label{fig:latentspace}
\end{figure}
Having identified the compression threshold at~\;$d=L-1$, we now examine the geometry of the learned latent representation at this critical dimension.
Fig.~\ref{fig:latentspace} illustrates feature-wise projections of the latent coordinates for test data (for a representative system size $L=N=6)$, color-coded by the strength of the external potential $\boldsymbol{\mu}$.
The potential $\boldsymbol{\mu}$ serves as the generative parameter used to sample distinct ground states; visualizing the latent encodings thus reveals how states at different potentials are represented.
As the mean potential $\bar{\mu} = \sum_j \mu_j$ shifts Eq.~\eqref{eq:energy} only by a constant and does not influence optimization, we quantify the strength of the potential by its standard deviation,
$\sigma(\boldsymbol{\mu}) = \|\boldsymbol{\mu} - \bar{\mu}\|_2$.
Color-coding by this quantity highlights how ground states with varying potential strengths are organized within the latent space.
The set of encoded latent vectors forms a nearly uniform sphere of radius $r \approx 2$.
This geometry arises from the well loss in Eq.~\eqref{eq:well} which confines embeddings to a bounded spherical region, while the contrastive repulsion in Eq.~\eqref{eq:repulsion} encourages uniform coverage.
Notably, the color map reveals a smooth radial gradient: 
points near the center correspond to weaker potentials, whereas those near the outer boundary represent stronger potentials.
The emergence of such structure implies that the encoding extracts physically meaningful information from the Hamiltonian terms $\boldsymbol{\omega}$.
Hence, the latent geometry demonstrates that the autoencoder learns a compact and physically organized representation of the ground state manifold.

While this visualization exposes correlations with the potential strength, it does not, by itself, reveal which physical quantities fundamentally shape the latent space.
In App.~\ref{appx:interpret}, we further analyze how the learned representation relates to the potential and the density, showing that it cannot be reduced to either descriptor alone.

\noindent \textbf{Variational Energy Optimization in Learned Latent Space} ---
Having characterized the geometry of the latent space, we now test the learned representation as a variational ansatz for ground-state search.
For a given test-set potential $\boldsymbol{\mu}$, we perform the minimization in Eq.~\eqref{eq:opt-latent} using L-BFGS~\cite{liu_limited_1989} with a learning rate of $8\times 10^{-2}$, starting from $\mathbf{z}_0 = \mathbf{0}$ and iterating until the convergence condition $\rVert\partial_\mathbf{z} E\rVert_2 \leq 3 \times 10^{-4}$ is met.

\begin{figure}[b]
    \centering
    \includegraphics[width=\columnwidth]{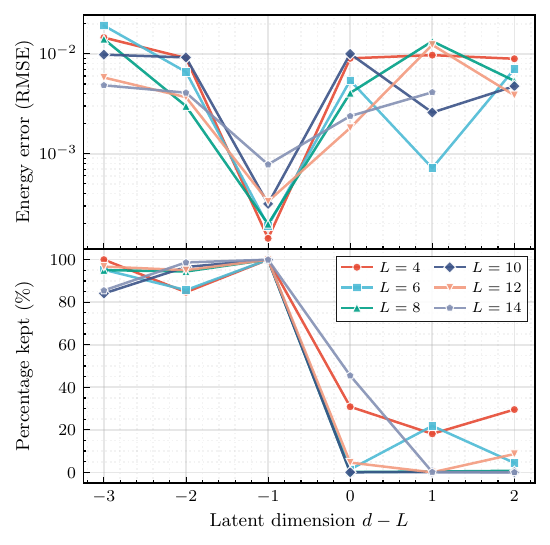}
    \caption{\textbf{Energy optimization threshold}:
        Error (RMSE) of optimized energies versus latent dimension $d$. Energies are obtained via gradient-based minimization in the learned latent space, using the decoder as a differentiable variational ansatz. A distinct error minimum occurs at $d = L - 1$, matching the intrinsic degrees of freedom of the half-filled Hubbard ground state. For $d \ge L$, performance degrades as excess latent dimensions introduce unconstrained, unphysical directions. Lower panel: Percentage of optimization trajectories retained after rejecting solutions violating the stability bound $\lVert \mathbf{z}^* \rVert < r_\text{opt}$ (note: for $L = 14$ at the largest $d$, all samples exceeded the bound). Results shown for even system sizes $L \in [4,\ldots,14]$ at $U/t = 4$.
    }
    \label{fig:en-opt}
\end{figure}

To avoid divergences $\lVert \mathbf{z}^* \rVert_2 \to \infty$ during optimization, we add an absorbing potential that enforces $\rVert \boldsymbol{z}\rVert_2 \leq r_\text{opt} = 3$.
This is implemented in two steps: 
we first augment the objective with a term analogous to Eq.~\eqref{eq:well}, but using a radius $r_\text{opt}$ larger than the training well; afterwards, we discard solutions with $\lVert \boldsymbol{z} \rVert_2 > r_\text{opt}$ and mark them as unreliable.
We repeat this procedure on $10^4$ test potentials.
Fig.~\ref{fig:en-opt} illustrates the root mean squared energy error of the accepted points, along with the fraction of rejected points, as a function of latent dimension $d$.

We observe that the error drops sharply as $d$ increases and reaches its minimum at $d = L - 1$, matching the compression threshold identified in Sec.~\ref{sec:results}.
At this dimension, the latent variables span the physically sufficient manifold supporting the Hubbard ground states.
For $d > L - 1$, the error increases again:
the additional latent directions introduce redundancy and weakly constrained, potentially unphysical degrees of freedom that deteriorate the conditioning of the variational landscape, even though reconstruction quality remains high.
In these regimes, the optimizer often pushes solutions toward the boundary of the allowed region, with $\|\mathbf z\|_2 \approx r_\text{opt}$, in an attempt to minimize the energy.

These results demonstrate that the autoencoder learns a latent representation that can serve as a practical variational manifold.
At $d = L-1$, the optimization landscape is empirically well conditioned and yields accurate ground-state energies without imposing explicit physical constraints.
For $d \ge L$, weakly constrained latent directions reduce stability, indicating that higher-dimensional latent spaces require additional regularization.

\newpage

\section{Discussion and Outlook}\label{sec:conclusion}
In this Letter, we have introduced an unsupervised machine-learning framework that learns compressed representations of fermionic quantum many-body ground states.
For the considered family of Hubbard Hamiltonians, we identify the minimal latent dimension $d=L-1$ required to learn a faithful representation.
This dimensionality matches the number of intrinsic degrees of freedom of the model.
For this choice of latent dimension, the representations of valid ground states form a smooth and bounded manifold reproducing the topology of the system's degrees of freedom (a $L-1$-dimensional ball).
The decoder can then serve as a stable and differentiable ansatz for valid ground states, enabling variational energy optimization.
Further increasing the latent dimension does not improve the quality of the learned representation. 
This, instead, introduces irrelevant directions in the latent space, orthogonal to the manifold of valid ground state representations, degrading the performance of the decoder as a variational ansatz.
Together, these results demonstrate that learned latent representations can serve both as probes of intrinsic dimensionality and as practical variational ansätze. 

The analysis of the encoder’s feature sensitivity (App.~\ref{appx:interpret}) indicates a strong reliance on local density and on-site terms, effectively rediscovering the importance of key variables central to Density Functional Theory. 
This establishes a natural connection to ML-assisted DFT approaches~\cite{luise_accurate_2025,kirkpatrick_pushing_2021,snyder_finding_2012,nelson_machine_2019} that leverage neural networks to refine exchange-correlation approximations. 

The Hubbard model offers an ideal starting point for our framework: 
its degrees of freedom are known a priori, providing a controlled testbed for benchmarking latent-space compression and interpretability.
Additional complexity could be introduced by extending the considered family of Hubbard Hamiltonians to include varying interaction ratios $U/t$; we expect this would require one additional latent dimension to capture the extra degree of freedom. 
However, because the parameter space is inherently anisotropic—i.e., changing the global $U/t$ impacts the underlying ground state distinctively compared to local potential shifts $\mu_i$—we anticipate that the resulting latent manifold will exhibit a more complex structure.
A natural next step is to apply this framework to systems where the relevant degrees of freedom are not analytically known, such as models of molecules and correlated materials, or results from quantum simulators. 
In such cases, the learned latent representation could act as a data-driven probe that reveals how many and which degrees of freedom actually matter for describing a family of ground states.

A variational ansatz learned from data could become useful if it can be applied to systems larger than those on which it was trained.
Future work may explore how such scalability can be achieved using geometric deep learning architectures, such as convolutional networks or transformers.
To extend our framework to more complex systems, the training pipeline could be adapted to rely on data generated by methods that are more efficient than exact diagonalization, such as tensor network techniques or quantum algorithms, and to incorporate datasets produced by a variety of computational approaches.
 
\section*{acknowledgments}
We thank Vladimir Zakharov, Anna Dawid, Björn van Zwol, Emanuele Costa, Evert van Nieuwenburg, and Vedran Dunjko for fruitful discussions.
This work was supported by the Dutch National Growth Fund (NGF), as part of the Quantum Delta NL program.
S.P.~and E.K.~acknowledge support from Shell Global Solutions BV.
The views and opinions expressed here are solely those of the authors and do not necessarily reflect those of the funding institutions.
Neither of the funding institutions can be held responsible for them.
The authors declare no competing interests.
Parts of this work were performed using the computational resources from the Academic Leiden Interdisciplinary Cluster Environment (ALICE) provided by Leiden University.

\section*{Code and Data Availability}
The data used for training the machine learning models, along with the code used to generate it and to reproduce the plots in this manuscript will be made available upon publication. 

\bibliography{zotero-export, manual-refs}

\appendix

\section{Dataset Details} \label{appx:dataset}
To generate training data for the neural-network-based autoencoder, we sample Hubbard model ground states at varying random potentials $\pot$ using the procedure detailed in Ref.~\cite{koridon_learning_2025} (an adaptation of the procedure from Ref.~\cite{nelson_machine_2019}), which we recap here. 

The first step in generating a data point is to sample a random potential $\mu$.
To do this, we first sample a strength parameter $W$ uniformly at random from the interval $[0.005t, 2.5t]$. We then sample the on-site potential $\pot$ uniformly at random within the range $[-W, W]$ and calculate its standard deviation $\sigma(\pot) = \sqrt{\langle \mu_j^2 \rangle - \langle \mu_j \rangle^2}$, where the average is taken over all sites $j$. 
If $\sigma(\pot) < 0.4t$, we accept the potential and proceed; otherwise, we reject it and repeat the sampling procedure from the beginning. 
This acceptance criterion produces a representative distribution of potentials with approximately uniformly-distributed standard deviations below $0.4t$, ensuring we sample both weak and strong potentials irrespective of the system size $L$.

Once a potential is accepted, we employ exact diagonalization to compute the ground state of the 1D Hubbard model at the given potential using the PySCF Full Configuration Interaction module~\cite{sun_pyscf_2018}. 
From the converged ground state $|\psi(\pot)\rangle$, we compute the two-particle reduced density matrix (2-RDM) 
\begin{equation}
    \Gamma_{pqrs}(\pot) := \sum_{\sigma \tau} \bra{\psi(\pot)} c^\dag_{p\sigma}c^\dag_{r\tau}c_{s\tau}c_{q\sigma}
    \ket{\psi(\pot)}. \label{eq:2rdm}
\end{equation}
We then extract the expectation values of Hamiltonian terms from the 2-RDM: 
the spin-summed on-site density $\langle\sum_\sigma\hat{n}_{i\sigma}\rangle$, the nearest-neighbor hopping correlators $\langle \sum_\sigma \hat c^\dag_{i\sigma} \hat c_{i+1,\sigma} \rangle$, and the on-site interaction $\langle\hat{n}_{i\uparrow}\hat{n}_{i\downarrow}\rangle $. 
This procedure is repeated for $N_{\text{inst}}=10^5$ independent potentials to generate the full dataset.
Stacking these three observables for all realizations yields a dataset of shape $(N_{\text{inst}}, 3 L)$.
Finally, by exploiting translational and reflection symmetries, we increase the effective size of the training dataset by a factor of $2 L$.
(this symmetry augmentation procedure is not applied to validation and test set).

\section{Neural Network Details}\label{appx:network}
To learn unsupervised compressions of the ground-state data, we employ an autoencoder with a symmetric encoder–decoder design. The input $\boldsymbol{\omega} \in \mathbb{R}^{n_{\text{in}}}$ contains the three local Hamiltonian terms across $L$ sites (thus, $n_\text{in} = 3L$). The encoder maps these observables to a latent vector $\mathbf{z}$ through four fully connected layers with Softplus activations, progressively reducing the dimensionality to a $d$-dimensional bottleneck: $E_{\boldsymbol{\theta}}: \boldsymbol{\omega} \mapsto \mathbf{z}.$
Here, $d$ is a tunable hyperparameter controlling the size of the latent space.
The decoder mirrors this structure, reconstructing the observables from the latent representation, $D_{\boldsymbol{\phi}}: \mathbf{z} \mapsto \boldsymbol{\omega}',$
using four hidden layers with Softplus activations that expand from dimension $d$ back to $n_{\text{in}}$. To keep the total parameter count largely independent of $d$, we impose a minimum hidden-layer width of $m_d = L^2$. 
The hidden-layer dimensions follow a geometric progression that interpolates smoothly between the narrow latent bottleneck and the wide reconstruction layers. 
Specifically, the first decoder layer starts at width $m_d$, and the final hidden layer terminates at $\lambda , n_{\text{in}}$, where the same scaling factor $\lambda = 20$ is also applied symmetrically to the encoder’s first hidden layer relative to the input dimension. 
This ensures a balanced expansion and contraction around the latent space.
This architecture provides a controlled and flexible framework for learning compact latent representations of the ground-state observables.

For the system sizes considered ($L = 4$ to $L = 14$), this architecture yields models with approximately $3\times 10^4$ to $7\times 10^5$ trainable parameters, scaling primarily with the input size.
For the compression of the two-body reduced density matrix, which has significantly higher input dimensionality, the largest model contains approximately $5\times 10^7$ parameters.
 
Training the largest models considered in this study empirically requires additional architectural modifications to prevent overfitting and ensure stable optimization. 
We implement this by soft-constraining the Lipschitz constant of the layers, which has the additional conceptual benefit of encouraging the model to learn a smoother mapping between latent space and feature space~\cite{liu_learning_2022}.
This is achieved with a modification of the fully-connected layers of the model and an additional term in the loss function.
To encourage Lipschitz smoothness, each layer $i$ applies row-wise weight normalization controlled by a learnable parameter $c_i$. 
For a weight matrix $W_i$, a bias $b_i$, and input $x$, the forward pass computes:
\begin{align}
    \hat W_i&=\mathrm{diag}\!\left(\min\!\Big[1,\tfrac{\mathrm{softplus}(c_i)}{\sum_k |(W_i)_{rk}|}\Big]_{r}\right) W_i,\\
    y&=\mathrm{Softplus}(\hat W_i x+b_i),
\end{align}
where the subscript $r$ indexes the rows of $W_i$. 
This normalization constrains the $\ell^\infty$ operator norm of each layer, with $\text{softplus}(c_i)$ serving as an adaptive, learnable upper bound on the maximum row sum.
The Lipschitz regularization loss in Eq.~\eqref{eq:lip} penalizes large Lipschitz bounds, encouraging the network to use the minimal Lipschitz constant necessary for accurate reconstruction. 
This term is added to the reconstruction loss during training with a hyperparameter weight $\gamma,\delta$, and the model is implemented using the normalization layers.

\section{Regularization Details}
We augment the reconstruction loss used to train neural network autoencoders with four regularization terms designed to yield a compact, smoothly structured latent manifold that meaningfully represents the ground state physics while remaining amenable to downstream optimization tasks.

To prevent unbounded latent encodings and ensure the learned representation occupies a compact region, we apply a radial well  potential as introduced in Eq.~\eqref{eq:well}, which penalizes latent vectors $\mathbf{z}$ that exceed radius $r$ from the origin. 
This soft confinement is essential for stable energy optimization in latent space, as it prevents the encoder from learning arbitrarily large embeddings that would complicate gradient-based search.

However, this confinement alone does not guarantee a well-structured latent space. 
Without additional constraints, the encoder could adopt a trivial solution in which all states collapse to a small region near the origin, satisfying the well loss while destroying any meaningful geometric structure. 
This phenomenon, known as latent collapse, causes distinct physical states to map to nearly identical latent codes, making the topology of the ground state manifold less accessible and rendering the learned representation useless for downstream tasks. 
To prevent this collapse and preserve the intrinsic geometry of the input space, we employ a contrastive repulsion loss~\cite{chen_simple_2020}.
For a batch $\{(\boldsymbol{\omega}_i, \mathbf{z}_i)\}_{i=1}^N$ of input states and their encodings, we penalize pairs that are well-separated in input space yet close in latent space, as introduced in Eq.~\eqref{eq:repulsion}.
This term encourages approximately isometric embeddings: 
pairs of states separated by a distance $d$ in input space should remain roughly a distance $d$ apart in latent space. 
The normalization by the average input distance makes the loss scale-invariant across different system sizes and potential strengths.

As introduced in App.~\ref{appx:network}, models trained on larger system sizes benefit from weight regularization.
Thus, to promote smooth latent-to-feature mappings and stabilize both training and subsequent latent-space optimization, we constrain the Lipschitz constant of the decoder network~\cite{liu_learning_2022}. 
For each decoder layer $\ell$ with weight matrix $W_\ell$, we introduce a learnable scalar parameter $c_\ell$ that bounds the layer's operator norm. 
Following the row-wise normalization scheme described in App.~\ref{appx:network}, we minimize the loss in Eq.~\eqref{eq:lip} which penalizes large Lipschitz bounds while allowing the network to adaptively learn the minimal smoothness required for accurate reconstruction. 
This regularization prevents exploding gradients during training and ensures that small perturbations in latent space correspond to small changes in the decoded output, a helpful property for gradient-based energy minimization.

The total training objective combines these terms:
\begin{equation}
    \mathcal{L} = \mathcal{L}_{\text{rec}} + \alpha \, \mathcal{L}_{\text{well}} + \beta \, \mathcal{L}_{\text{repel}} + \gamma \mathcal{L}_{\text{lip}}^{\text{enc}}+ \delta \mathcal{L}_{\text{lip}}^{\text{dec}},
\end{equation}
where we set $\alpha = 10^{-7}$, $\beta = 10^{-7}$, $\gamma = 10^{-9}$, and $\delta = 10^{-8}$ across all experiments. 
These coefficients balance the regularization effects in relation to reconstruction accuracy.

\section{Training and Hyperparameter Details}
We implement all models in PyTorch~\cite{paszke_pytorch_2019}, train them on GPU using the Adam optimizer~\cite{kingma_adam_2017} with a batch size of $256$, and use a $81/9/10$ train–validation–test split.
Training runs for up to $3\times 10^3$ epochs with early stopping: 
if the validation loss fails to improve by at least $\Delta = 10^{-10}$ for $30$ consecutive epochs, training terminates. 

The learning rate schedule begins at $\eta_0 = 4\times 10^{-3}$ and adapts using PyTorch's ReduceLROnPlateau scheduler, which reduces $\eta$ by a factor of $0.5$ whenever the validation loss plateaus (improvement $< 10^{-8}$) for more than $10$ epochs. 
This adaptive schedule allows the optimizer to make aggressive updates early in training while refining the solution with smaller steps as convergence approaches.

Hyperparameters, including the initial learning rate, batch size, number of hidden layers $D$, width scaling factor $\lambda$, minimum layer width $m_d$, and regularization coefficients $\{\alpha, \beta, \gamma, \delta\}$, are selected with the guidance of the Optuna framework~\cite{akiba_optuna_2019}. 
The best-performing configuration from this search is then used to train the final model reported in all results.

All reported performance metrics are evaluated on a held-out test set of potential configurations $\{\boldsymbol{\mu}\}$ that were not seen during training or hyperparameter tuning.

\section{Ablation Study}\label{apx:ablation}
\begin{figure}[b]
    \centering
    \includegraphics[width=\columnwidth]{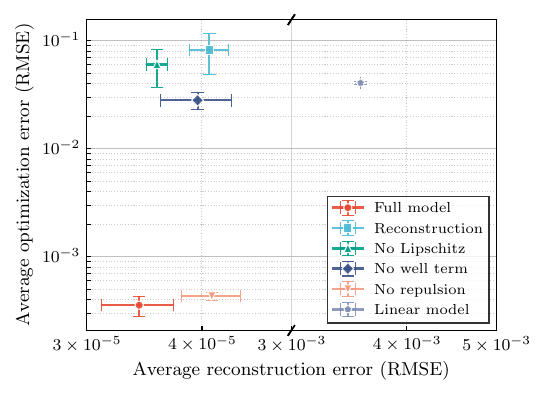}
    \caption{
    \textbf{Ablation study}:
    Performance comparison of model variants on systems with $L = N = 6$ and latent dimensions $d = L - 1$. 
    We evaluate six distinct model configurations in terms of reconstruction error and optimization performance. 
    Error bars show the standard deviation over five independently trained models with different random initializations. 
    The full model consistently outperforms all ablated variants, indicating that each regularization component contributes meaningfully to both accurate reconstruction and effective energy optimization.
    }
    \label{fig:ablation}
\end{figure}

To understand the contribution of each architectural and regularization component to the model's performance, we conduct a systematic ablation study on the $L=N=6$ system. 
We compare six model configurations, evaluating both reconstruction accuracy and optimization performance. 
For each configuration, we train five independent models with different random initializations to quantify variability, reporting mean and standard deviation across these runs.
We evaluate the following model variants:

\begin{itemize}
    \item \textbf{Full model}: 
    Nonlinear encoder and decoder with $D=4$ hidden layers, trained with all regularization terms ($\mathcal{L}_{\text{rec}} + \alpha \, \mathcal{L}_{\text{well}} + \beta \, \mathcal{L}_{\text{repel}} + \gamma \mathcal{L}_{\text{lip}}^{\text{enc}}+ \delta \mathcal{L}_{\text{lip}}^{\text{dec}}$).
    
    \item \textbf{Reconstruction only}: Full architecture but trained with $\mathcal{L}_{\text{rec}}$ alone, removing all regularization ($\alpha = \beta = \gamma = \delta= 0$). 
    This tests whether regularization provides benefits beyond accurate reconstruction.
    
    \item \textbf{No Lipschitz regularization}: 
    Trained without smoothness constraints ($\gamma =\delta = 0$), while retaining well confinement and contrastive repulsion. 
    This isolates the effect of enforcing smooth latent-to-output mappings.
    
    \item \textbf{No well loss}: Trained without confinement ($\alpha = 0$), allowing unbounded latent encodings. 
    This tests whether compact latent regions are necessary for effective optimization.
    
    \item \textbf{No contrastive repulsion}: 
    Trained without the anti-collapse term ($\beta = 0$). 
    This examines whether preventing latent collapse is critical for preserving the ground state manifold geometry.
    
    \item \textbf{Linear model}:
    Replaces the nonlinear layers in encoder and decoder with a single linear layer. 
    This tests whether nonlinearity in the model is essential.
\end{itemize}

All models use identical latent dimensions $d=L-1$, training procedures, and hyperparameters (excluding the ablated components) to ensure a fair comparison.
We evaluate each model configuration using two complementary metrics.
First, we measure the reconstruction error on a held-out test set, which quantifies how faithfully the autoencoder can compress and reconstruct the Hamiltonian-term features.
Second, to assess how well the learned latent space supports energy minimization, we perform gradient-based optimization in latent space to determine ground states for previously unseen potential configurations.

Fig.~\ref{fig:ablation} summarizes reconstruction error versus optimization error for all model variants, with error bars showing the standard deviation across five random initializations.
The results reveal several key observations.

\textbf{Reconstruction Error} ---
All nonlinear models, independent of which regularization terms were enabled, achieve reconstruction errors of comparable magnitude.
This shows that the nonlinear encoder–decoder architecture is generally capable of faithfully representing the Hamiltonian-term features within the latent bottleneck of size $d = L - 1$.
Among these, the full model achieves the best reconstruction performance, indicating that each regularization component contributes positively.
In contrast, the linear model performs two orders of magnitude worse, confirming that it lacks the expressivity required for the task.
\begin{figure}[b]
    \centering
    \includegraphics[width=\columnwidth]{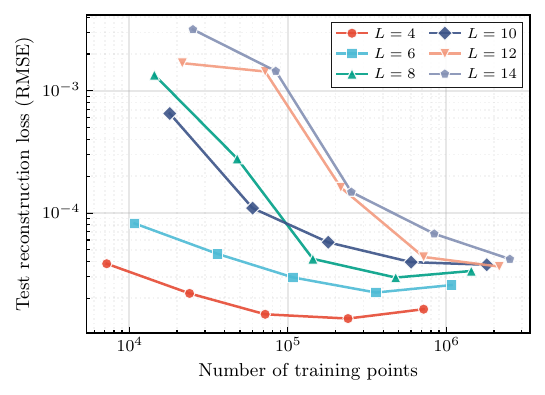}
    \caption{
        \textbf{Training data efficiency}: 
        root mean squared error reconstruction loss as a function of the size of the training and validation dataset (accounting for symmetry augmentation) for different system sizes at the optimal latent dimension $d = L - 1$.
        The curves indicate how the data requirement grows with system size to achieve high-quality compression and display the best out of three training instances.
    }
    \label{fig:training-data-efficiency}
\end{figure}

\textbf{Optimization Error} ---
The full model provides the best overall trade-off, achieving the lowest optimization error.
In contrast, the reconstruction-only model performs the worst.
Although it achieves reasonable reconstruction accuracy, its unregularized and unbounded latent space leads to poor optimization behavior.
The model without Lipschitz regularization maintains good reconstruction performance but also shows degraded optimization behavior.
From this, we can infer that encouraging Lipschitz smoothness improves latent-to-output geometry: 
small steps in latent space correspond to controlled, predictable changes in the reconstructed observables, which is essential for stable and reliable minimization.
The model without the well constraint suffers from a similar issue.
Without this constraint, the latent space can expand in an unbounded or highly non-uniform manner due to the repulsion term, making the gradient-based search for the minimum energy substantially harder.
The model without the repulsion term shows solid optimization performance but slightly reduced reconstruction quality.
This suggests that repulsion primarily helps structure the latent space in a way that improves reconstruction with only a mild influence on optimization when the other regularizers are still present.
Interestingly, despite its poor reconstruction, the linear model achieves an optimization error comparable to the reconstruction-only baseline.
We attribute this to its convex and well-behaved latent geometry, which facilitates gradient-based minimization even though the representation is less optimal for accurate reconstruction.

Overall, these ablations illustrate how each architectural component contributes to a latent space that is both expressive enough for accurate reconstruction and structured enough to support robust energy optimization.

\section{Training Data Efficiency}\label{appx:training-data}
In the main experiments, the autoencoder was trained on $2L \times 10^5$ Hamiltonian term expectation values $\boldsymbol{\omega}$, generated by $10^5$ unique potentials $\pot$.
A natural question is how the amount of training data required to learn optimal compression scales with the system size $L$.
Fig.~\ref{fig:training-data-efficiency} illustrates the relationship between the size of the training and validation set (already accounting for the $2 L \times$ increase from the symmetry augmentation described in App.~\ref{appx:dataset}) and the reconstruction loss for various system sizes, evaluated at the critical latent dimension $d = L - 1$.
The point at which the reconstruction error drops sharply provides an approximate measure of the data required for efficient learning at this compression threshold.
We observe a marginal increase in test reconstruction error for smaller system sizes when using the largest dataset sizes. 
We attribute this minor fluctuation to both the fixed 
hyperparameter configuration, which was optimized for the baseline dataset size of $10^5$ points, and to the training dynamics of the smaller model architecture (approx. $3\times10^4$ parameters for $L=6$). 
Crucially, this discrepancy remains well below the convergence tolerance of $3\times10^{-4}$ (RMSE) used for variational energy optimization.

\section{Interpreting the Encoder}\label{appx:interpret}
While the visualization in Fig.~\ref{fig:latentspace} demonstrates that the learned representation is compact and varies smoothly with potential strength, it does not reveal which physical features fundamentally structure the latent space.

Interpreting deep nonlinear encoders is inherently challenging. 
Unlike in linear models, individual weights are not directly meaningful, and the effect of any input feature can vary substantially across different regions of the input space. 
In machine learning research, this challenge is often addressed using gradient-based saliency maps~\cite{smilkov_smoothgrad_2017,simonyan_deep_2014,sundararajanTY17} or SHAP-style local attribution methods~\cite{rozemberczki_shapley_2022,lundberg_unified_2017}, which quantify how sensitive a model’s output is to perturbations of each input feature.

\begin{figure}[h!]
    \centering
    \includegraphics[width=\columnwidth]{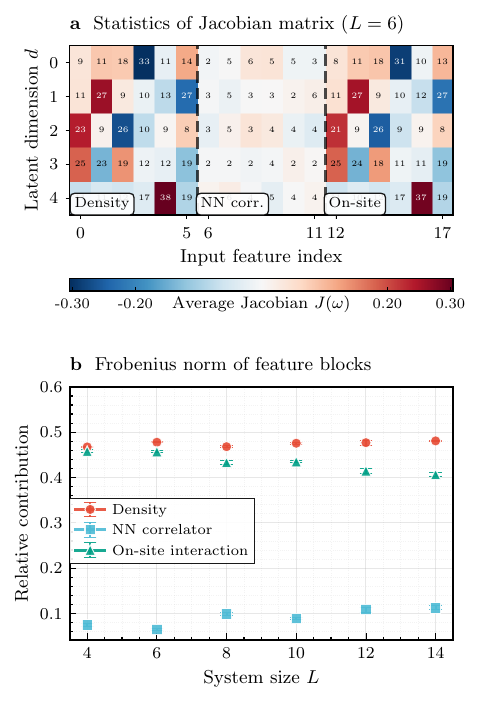}
    \caption{
    \textbf{Interpreting the trained encoder}:
    (a) Average Jacobian matrix and variability. 
    Visualization of the mean Jacobian matrix $J(\boldsymbol{\omega})$ of the encoder for the $L=N=6$ model (latent dimension $d=L-1$) computed over $10^4$ input samples. 
    The color map represents the mean magnitude ($\mu$), while the overlaid numbers represent the sample standard deviation ($\sigma$), scaled by a factor of $10^{3}$. 
    The consistently small variability ($\sum \sigma_{i,j} / \sum |\mu_{i,j}|\approx16\%$
    ) across the dataset suggests that the encoder has learned a stable, close to linear mapping from the input feature space to the latent representation. 
    Distinct patterns are visible across the three input feature blocks: 
    spin-summed density and on-site interaction exhibit larger average gradients, while the nearest-neighbor pair correlator appears less influential.
    (b) Relative feature importance via frobenius norm. 
    Due to the inherent correlation between input features and the linear dependencies within the latent space, the element-wise Jacobian can only offer limited insight into feature importance. 
    To provide a robust, rotationally invariant measure of effective sensitivity, we calculate the relative Frobenius norm contribution of each feature's Jacobian submatrix~\eqref{eq:relfrob} as a function of system size $L$ (at fixed latent dimension $d=L-1$). 
    The error bars shown are averaged over three independently trained models. 
    This heuristic confirms a consistent trend across system sizes: 
    the spin-summed density and on-site interaction are the dominant features encoding information into the latent representation, while the nearest-neighbor correlator remains the least significant.
    }
    \label{fig:weight_analysis}
\end{figure}
\clearpage

For our encoder, the natural analog of such attribution tools is the Jacobian: \begin{equation}
    J(\boldsymbol{\omega})= \partial E_{\boldsymbol{\theta}}( \boldsymbol{\omega})/\partial \boldsymbol{\omega}
\end{equation}
This matrix measures how each latent coordinate responds to infinitesimal changes in each input observable at a specific sample, thus providing a precise notion of local feature relevance for nonlinear mappings. 
The main limitation of this approach is the same locality: 
if the encoder were strongly nonlinear, the Jacobian would vary substantially across input samples, making a single global interpretation unreliable.

To assess the feasibility of a global interpretation, we compute the Jacobian for a representative set of $10^4$ ground-state data points for the $L = 6, \ d = L - 1$ model. 
We then analyze the mean and standard deviation of each Jacobian element (Fig. $\ref{fig:weight_analysis}$a) . 
In this visualization, the color map represents the mean magnitude ($\mu$), and the overlaid numbers denote the sample standard deviation ($\sigma$), which is scaled by a factor of $10^{3}$ for improved readability.

Two key observations emerge from this analysis:
First, the Jacobians vary only weakly across the dataset. 
The total relative variability $\sum \sigma_{i,j} / \sum |\mu_{i,j}|\approx16\%$ across the dataset is small, indicating that the encoder applies nearly the same linearized transformation throughout the data manifold. 
This near-constancy implies that the encoder is close to a linear function, and the Jacobian can be interpreted almost like the weight matrix of a linear encoder.
Second, the average Jacobian magnitude clearly shows three distinct blocks corresponding to spin-summed density, on-site interaction, and nearest-neighbor correlators.
Density and on-site terms exhibit consistently larger magnitudes. 
This means that perturbations in these observables produce stronger change in the latent representation. 
In contrast, the nearest-neighbor correlator block has substantially smaller magnitudes, indicating that these features play a comparatively minor role in shaping the learned latent coordinates.

In summary, because the encoder operates in an effectively linear regime over the data manifold, the Jacobian magnitudes provide an interpretable measure of feature importance. 
The observed block structure highlights which physical observables dominate the latent representation and gives insight into how the encoder processes the input features.

To obtain a robust characterization of feature influence across different system sizes, we evaluate the relative Frobenius norm contribution of each feature's Jacobian submatrix. This metric, $r_f$, is defined as:

\begin{equation}
r_f = \frac{||J_f(\boldsymbol{\omega})||_F}{\sum_{f'} ||J_{f'}(\boldsymbol{\omega})||_F}, \label{eq:relfrob} 
\end{equation}
where $J_f(\boldsymbol{\omega})$ denotes the block of the Jacobian matrix associated with feature $f$. 
The Frobenius norm $||J_f(\boldsymbol{\omega})||_F = \sqrt{\sum_{i,j} (J_f^{ij}(\boldsymbol{\omega}))^2}$ serves as a natural measure of the total sensitivity of the latent representation to changes in feature $f$, aggregating information across all latent dimensions and spatial positions.

The Frobenius norm captures the overall strength of the feature's influence, preventing cancellation that occurs when averaging raw, signed gradients and is invariant to orthogonal transformations of the latent basis. 
The relative normalization ensures that contributions sum to unity, enabling direct comparison of feature importance across different system sizes and architectural choices.

Figure $\ref{fig:weight_analysis}$(b) shows these relative Frobenius norm contributions as a function of system size for $d = L - 1$, with the mean and standard deviation over three independently trained models.  
A consistent trend is observed: density and on-site interaction terms dominate the learned representation across all sizes, while the contribution of the nearest-neighbor correlator remains low. 
This pattern suggests that the autoencoder primarily relies on local quantities that directly encode the potential and interaction structure of the Hamiltonian, with longer-range correlations only weakly influencing the compact latent description. 
The narrow error bars confirm that this hierarchy of importance is robust.

\section{Degeneracies in ground state manifold}\label{appx:odd}
In Sec.~\ref{sec:results}, we studied system sizes $L \in \{4, \ldots, 14\}$ with even particle numbers at half-filling. 
To explore the limits of our compression approach, we also test our framework on systems with an odd number of sites $L$, close to half filling $N=L\pm1$.
In particular, we report the results for systems with $(L=5, N=4)$ and $(L=7, N=8)$.
We identified these systems as exhibiting different behavior from others during training. 
In particular, they achieve a slightly worse reconstruction loss ($\approx 1e-4$) compared to other systems of a similar size ($\approx 3e-5$) for $d\geq L-1$.

\begin{figure}[!t]
    \centering
    \includegraphics[width=\columnwidth]{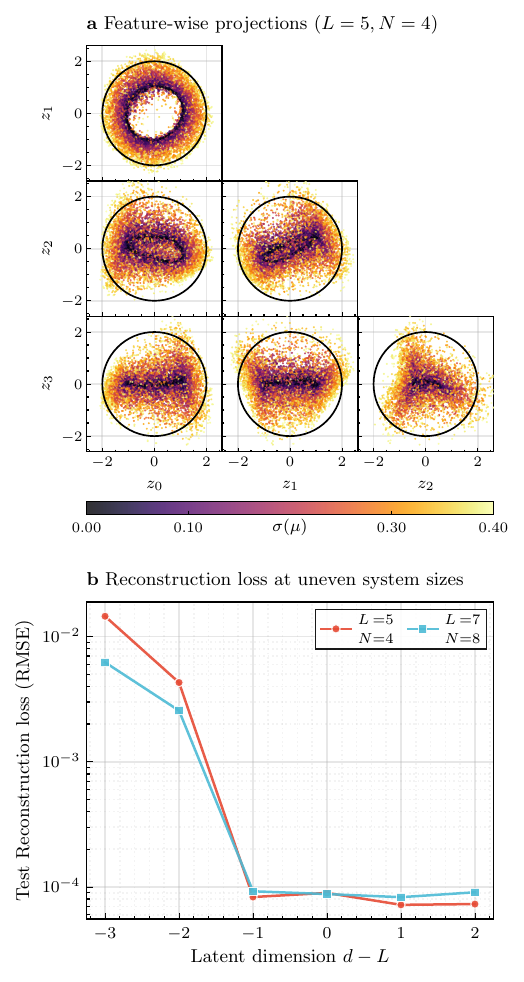}
    \caption{\textbf{Compression of degenerate systems}:
    (a) Latent space of the autoencoder for $(L = 5, N = 4)$, color-coded by the potential strength $\boldsymbol{\mu}$.
    Unlike the $L=N$ systems, the data does not exhibit the same circular structure in the feature-wise projections, indicating qualitatively different underlying physics.
    The ring-like structure illustrates an underlying degeneracy in the ground state manifold.
    (b) Test reconstruction loss as a function of latent dimension $d$.
    A sharp decrease in reconstruction error is again observed at $d = L - 1$, but the overall reconstruction quality is comparatively worse than for the even systems.
    }
    \label{fig:odd_even}
\end{figure}

For these systems,
Fig.~\ref{fig:odd_even}(a) illustrates the feature-wise projections of the latent space at the critical dimension $d = L-1$ and (b) the reconstruction loss as a function of the latent dimension.
The results reveal markedly different behavior compared to the even-$L$ systems studied previously.
We observe that the encoded points extend well beyond the radial well boundary $r$, indicating that the confinement regularization is less effective. 
Moreover, the distribution is highly anisotropic. 
The feature projections show that the topology of the representation manifold is that of a ring, rather than the spherical one observed for even-$L$ systems (see Fig.~\ref{fig:latentspace}).

This qualitative difference reflects an underlying quasi-degeneracy in the spectrum of the quantum system. 
In Fermi-Hubbard chains with periodic boundary conditions, with $N = 4k$ (where $k \in \mathbb{N}$) and odd $L$,
the singlet ground state of the zero-potential system $\vec\mu = 0$ is doubly degenerate.
(This can be observed both numerically and explained by analyzing the model Hamiltonian in momentum space. The degenerate ground states can be labeled by the total momentum, which is a good quantum number when $\boldsymbol{\mu} = 0$.) 
We define $\ket{\phi_\text{A}}$ and $\ket{\phi_\text{B}}$ as two orthonormal vectors forming a basis for the ground subspace of $H(\boldsymbol\mu=0)$.
While a non-zero disorder potential $\boldsymbol{\mu}$ lifts this degeneracy, the energy splitting is perturbatively small for potentials $\boldsymbol{\delta\mu}$ with weak fluctuations $\sigma(\boldsymbol{\delta\mu}) \ll t$.
This symmetry breaking generates arbitrary combinations of the two degenerate zero-potential states for perturbatively small potentials; thus, the ground state of $H(\boldsymbol{\delta\mu})$ can be written as $a \ket{\phi_\text{A}} + b \ket{\phi_\text{B}} + o(
\sigma(
\boldsymbol{\delta\mu})
)
$.
As the considered model Hamiltonians can be represented as real symmetric matrices, the coefficients $a,b \in \mathbb{R}$; equivalently $a = \sin{\theta}, b=\cos{\theta}$.
Thus, the manifold of ground states for infinitesimal potentials has a $\textsc{SO}(2)$ structure (i.e.~a circle).
The latent representation faithfully captures this structure, with 
the ring topology visible in Fig.~\ref{fig:latentspace}.
The inner edge of the ring corresponds to the ground states of models with a weakly-fluctuating potential.
This is precisely the behavior we expect for a system exhibiting disorder-induced symmetry breaking of a doubly-degenerate ground state.

This analysis demonstrates that reconstruction quality, as well as latent space visualization, serves as a diagnostic for ground state complexity beyond simple compressibility. 
Systems with accidental or near-degeneracies present a fundamentally harder learning problem: 
the autoencoder must either (i) commit to one branch of the degenerate manifold, breaking the underlying symmetry, or (ii) represent the full degenerate subspace, which require additional latent dimensions or nonlinear manifold structures (such as the observed rings).
The even-$L$ systems studied in the main text avoid this issue due to the absence of such degeneracies at generic disorder strengths, enabling clean, low-dimensional compressed representations.

\section{Compression of Two-Body Reduced Density Matrices} \label{app:2rdm}
While the Hamiltonian terms $\boldsymbol{\omega}$ provide a compact representation sufficient for energy evaluation, they capture only limited information about ground state correlations. 
A strictly more informative representation is the two-particle reduced density matrix introduced in Eq.~\eqref{eq:2rdm}, which encodes all two-body correlation functions of the ground state and finds application in various quantum chemistry settings~\cite{coleman_structure_1963,lowdin1955quantum}.

\begin{figure}[t!]
    \centering
\includegraphics[width=\columnwidth]{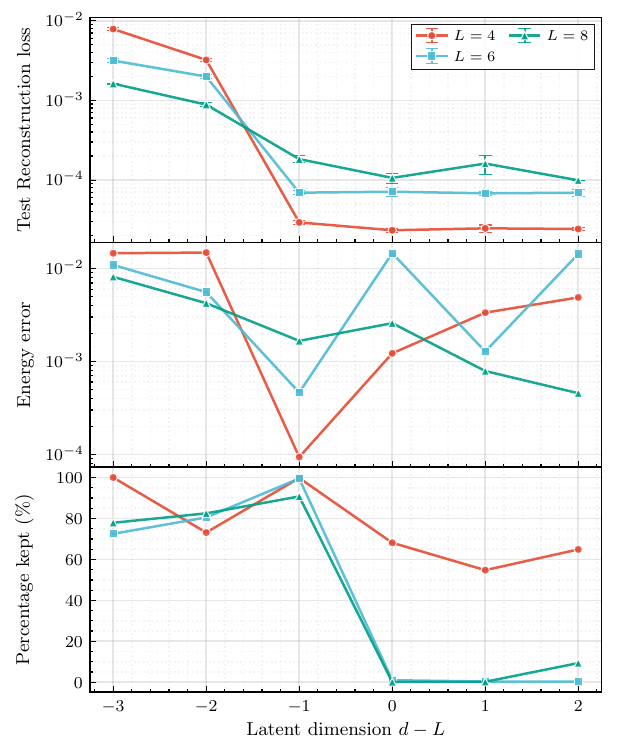}
    \caption{ 
    \textbf{Compression and energy optimization of two-body reduced density matrices}:
    (Top) Test reconstruction RMSE (mean and standard deviation over three models) versus latent dimension $d$. A sharp drop at $d = L - 1$ confirms that the 2-RDM shares the same intrinsic degrees of freedom as the Hamiltonian terms, despite its larger dimensionality.
    (Middle) Energy optimization error using the decoder as a variational ansatz. The critical dimension $d = L - 1$ yields the optimal performance; for $d \ge L$, the unconstrained latent directions degrade the optimization stability.
    (Bottom) Fraction of optimization trajectories retained after rejecting solutions that violate the stability bound $\lVert \mathbf{z}^* \rVert < r_\text{opt}$.
    Data shown for system sizes $L \in [4, 6,8]$ at $U/t = 4$.
    }
    \label{fig:2rdm_compression}
\end{figure}

We now apply the same unsupervised autoencoding framework to learn compressed representations of 2-RDMs instead of Hamiltonian terms.
Since the 2-RDM is a substantially larger object with dimension $L^4$, we are limited to system sizes $L \leq 8$ due to memory and training constraints.
We use $10^5$ samples without symmetry augmentation to train, validate, and test the models.
Fig.~\ref{fig:2rdm_compression}(top) shows the reconstruction error versus latent dimension for 2-RDM autoencoders on systems $L \in [4, 6, 8]$. 
We observe qualitatively similar compression behavior to Hamiltonian terms (Fig.~\ref{fig:compression}): 
a critical dimension emerges at $d = L-1$, below which reconstruction error does not decrease significantly.
We observe a minor deviation for the $L=8$ system, which we attribute to the limited training data available relative to the high dimensionality of the 2-RDM feature space. 
Consistent with the data efficiency analysis in Fig.~\ref{fig:training-data-efficiency}, we anticipate that 2-RDM models would benefit from expanding the effective dataset size via symmetry augmentation. 
We leave the detailed exploration of this scaling regime for future study.

Despite the 2-RDM's  larger input dimensionality, the optimal latent dimension remains $d=L-1$, identical to that of the Hamiltonian terms expectation values. 
This again confirms that the intrinsic dimensionality of the ground state manifold is determined by the underlying physical degrees of freedom and not by the representation's size. 
Since the Hamiltonian terms can be extracted as linear contractions of the 2-RDM and yield the same compression threshold, we conclude that the higher-order correlation information encoded in the 2-RDM does not add independent degrees of freedom for the Hubbard ground states studied here. 
The two-body correlations beyond those captured in local observables lie in a redundant subspace that does not require additional latent dimensions to represent.

We perform the same latent-space energy optimization introduced in Sec.~\ref{sec:results} using a trained 2-RDM autoencoder. 
Fig.~\ref{fig:2rdm_compression} (middle and bottom panels) demonstrates that the optimization achieves qualitatively similar performance to Hamiltonian term compression, tested for $800$ trial potentials.
We attribute this to the high reconstruction fidelity achieved during training; 
the autoencoder learns to stay within the physically accessible subspace of the 2-RDM space, implicitly satisfying the $N$-representability conditions~\cite{mazziotti_structure_2012,mazziotti_two-electron_2012} without explicit enforcement.
Fig.~\ref{fig:2rdm_compression}(bottom) again illustrates that for latent dimensions $d > L - 1$, the optimizer pushes solutions toward the boundary of the allowed region, with $\|\mathbf z\|_2 \approx r_\text{opt}$, in an attempt to minimize the energy, which results in an increasing number of rejected results.

\clearpage

\end{document}